\documentclass[twocolumn,showpacs,aps,%
prb,amsmath,amssymb,floatfix,superscriptaddress]{revtex4}

\usepackage{graphicx} 
\usepackage{dcolumn}
\usepackage{bm}
\begin{document}
\title{Centipede ladder at quarter filling }
\author{D.N. Aristov}
\altaffiliation[On leave from ] {Petersburg Nuclear Physics
Institute, Gatchina  188300, Russia} \affiliation{Institut f\"ur
Theorie der Kondensierten Materie, Universit\"at Karlsruhe,  Germany
} \affiliation{Center for Functional Nanostructures, Universit\"at
Karlsruhe, 76128 Karlsruhe, Germany}
\author{M.N. Kiselev}
\affiliation{The Abdus Salam Internetional Centre for Theoretical
Physics, Strada Costiera 11, Trieste, Italy}
\author{K. Kikoin}
\affiliation{School of Physics and Astronomy, Tel-Aviv University,
Tel-Aviv 69978, Israel}
\date{\today}
\begin{abstract}
We study the ground state and excitation spectrum of a quasi
one-dimensional nanostructure consisting of a pole and rungs
oriented in the opposite directions ("centipede ladder", CL) at
quarter filling. The spin and charge excitation spectra are found in
the limits of small and  large longitudinal hopping $t_\|$ compared
to the on-rung hopping rate $t_\perp$ and exchange coupling
$I_\perp$. At small $t_\|$ the system with ferromagnetic on-rung
exchange demonstrates instability against dimerization. Coherent
propagation of charge transfer excitons is possible in this limit.
At large $t_\|$ CL behaves like two-orbital Hubbard chain, but the
gap opens in the charge excitation spectrum thus reducing the
symmetry from $SU(4)$ to $SU(2)$. The spin excitations are always
gapless and their dispersion changes from quadratic magnon-like for
ferromagnetic on-rung exchange to linear spinon-like for
antiferromagnetic on-rung exchange in weak longitudinal hopping
limit.
\end{abstract}

\maketitle
\section*{INTRODUCTION}
The spin ladder systems attract much interest in both experimental
and theoretical communities \cite{dgrice} during recent years. On
the one hand, the ladder systems may be viewed as  intermediate
configuration between purely one-dimensional systems and periodic
arrays of higher dimensions. On the other hand, in many cases the
simple theoretical models represent a good prototype for description
of fundamental phenomena like metal-insulator transition, formation
of various density wave states, low dimensional superconductivity
and many other strongly correlated effects \cite{dgrice}. Besides,
some of the toy theoretical models are simple enough to be solved
exactly on the lattice or in the long wave continuum limit
\cite{GNT,ess}. The important experimental realizations of the
ladder systems (see e.g. \cite{dgrice}) renewed an interest to
imperfect spin chains and ladders affected by the chemical
substitution, pressure or radiation defects. Although the literature
on symmetric two-leg ladders at half-filling is quite extensive
\cite{dgrice, GNT, giamarchi}, not so much is known about the
effects of strong asymmetry in the exchange interaction (or hopping)
between the legs \cite{SNT}. Even less is known about the doped
two-leg ladders away from the half-filling regime \cite{ctrtsv,
exper1}. The most of considered cases are related to so-called
incommensurate filling where in spite of strong Coulomb (Hubbard)
interaction, the ground state of the ladder is metallic. On the
contrary, at the commensurate filling (e.g. quarter filling) the
Umklapp processes open a gap in charge excitations spectrum already
for weak interaction. These effects can be consistently treated by
means of bosonization technique \cite{GNT,giamarchi}.

\begin{figure}
\includegraphics[width=0.45\textwidth]{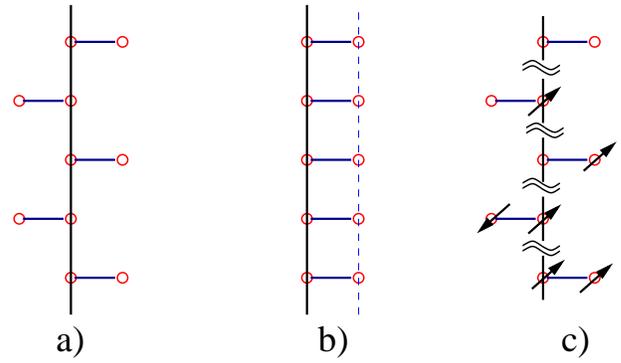}\\
\caption{(Color online) \label{fig:f1} Centipede (a) and strongly
asymmetric two-leg ladder (b). Several possible rung states (c)
corresponding to vacancy, single occupied states, singlet and
triplet states. Excited states with double on-site occupancy are not
shown.}
\end{figure}

In this paper we consider a model characterized by a gap opened due
to kinematic constraints even before any interactions are taken into
account. This model, which we call a centipede ladder (CL) or
single-pole ladder (see Fig.1a,c), can be visualized as the strongly
asymmetric limit of the two-leg ladder (Fig.1b) where the
interaction along one of the two legs is negligibly small compared
to both interaction along the main leg and the rung, e.g. due to
spatial orientation of the rungs (see Fig. \ref{fig:f1}a,b).

The family of centipede ladder-type systems schematically
presented in Fig. \ref{fig:f1}a was synthesized as a stable
organic biradical crystal PNNNO \cite{exp}. The analysis of this
geometry for the case of half-filling (two electrons on the rung
with infinite on-site repulsion)  has shown \cite{kak05a,kak05b}
that this model differs from the symmetric two-leg ladder. The
half-filled case for the centipede ladder belongs to the specific
class of universality manifesting itself in fractional (2/3)
scaling of the spin gap. This new scaling was attributed to
existence of hidden dynamical symmetries associated with quantum
nature of spin-rotators formed at each rung of the ladder and
should be contrasted with the integer scaling discovered for the
symmetric two-leg ladders \cite{SNT}. The corresponding model of
highly asymmetric two-leg ladder where the interaction between the
spins residing at the end of neighboring rungs is zero has been
labeled as Spin Rotator Chain model, \cite{kak05a} because it may
be mapped onto an effective 1D chain with complicated on-site and
inter-site exchange interaction revealing "hidden" symmetries of
spin rotator.

 One may expect that the same geometry of centipede
ladder away from the half-filling may also demonstrate a behavior
very different from such for symmetric two-leg ladder and/or the
partially filled chains.  A quarter-filled CL is the next
commensurate system where these differences should be looked for.
At first sight, the quarter-filled CL, with strictly one electron
per rung is equivalent to the quarter-filled two-orbital Hubbard
chain (HC)\cite{Yama} with partially lifted orbital degeneracy.
However, additional local symmetries related to the spin and
charge degrees of freedom of electrons on a rung influence a
structure of the ground state and the spectrum of low-energy
excitation and makes the phase diagram of CL more reach than that
of half-filled HC or quarter-filled two-orbital HC. The aim of
this paper is to reveal these new features. The phase diagram of
CL promises to be complicated enough, and in this paper we confine
ourselves with several limiting cases.

The manuscript is organized as follows: in  Section I we formulate
the model and describe the hierarchy of the electronic states on the
rung.   Section II is devoted to analysis of the limit, where the
on-rung ferromagnetic exchange is dominant. In the Section III we
consider the case of weak longitudinal hopping for antiferromagnetic
on-rung exchange. The Section IV is devoted to the case of strong
longitudinal hopping and discussion of various response functions at
quarter filling. Details of calculations are presented in Appendix.
In Conclusions the summary of results and the perspectives is
discussed.
\section{Model and hierarchy of rung states}
The generic model Hamiltonian is ${\cal H}={\cal H}_r+{\cal H}_l$,
where
     \begin{eqnarray}\label{Hr}
     && {\cal H}_r =
     \epsilon_0\sum_{si\sigma}n_{si,\sigma} -
t_\perp\sum_i\left(
        c^\dag_{\alpha i,\sigma}c_{\beta i,\sigma}+ {\rm H.c}\right)\\
&&     +\frac{1}{2}
     \sum_{si,\sigma} U
     n_{si,\sigma}n_{si,\bar\sigma}
   -  I_\perp\sum_i\left({\bf s}_{\alpha i}{\bf
        s}_{\beta i}-\frac{1}{4}n_{\alpha i}n_{\beta i}\right)\label{hm}, \nonumber
\end{eqnarray}
describes the electronic states on the rungs, and
\begin{eqnarray}\label{Hl}
{\cal H}_l= -\left(t_\parallel\sum_{i\sigma}
        c^\dag_{\alpha i,\sigma}c_{\alpha i+1,\sigma}+ {\rm H.c.}\right)
\end{eqnarray}
is related to the longitudinal electron hopping along the leg.
Here $s=\alpha,\beta$ enumerates sites belonging to the leg
$(\alpha)$ and the branch $(\beta)$ in a given rung $i$. The
Hamiltonian (\ref{hm}) is is an extension of the Spin Rotator
Chain Hamiltonian \cite{kak05a, kak05b} to the case where empty
states on the rungs appear and electron hopping is possible both
within a given rung $(t_\perp)$ and between neighboring rungs
$(t_\parallel)$. We retained here the direct on-rung exchange
interaction $I_\perp$, which can be either ferromagnetic or
antiferromagnetic.

It is natural to use a single rung basis for characterization of the
energy spectrum of SRC. The set $\mathbb{S}$ of 16 rung basis states
consists of an empty state $|0\rangle$ (which we will call a
vacancy), four singly occupied rung states, four doubly occupied
rung states which are split into singlet/triplet configurations, two
doubly occupied on-site states, four triply occupied states and a
 fully occupied state with four electrons. In the limit of strong Hubbard repulsion,
 $U\gg t_\|$, the lowest  states on the rung $i$ belonging to the
 charge sector $N_i=2$ (doubly occupied rungs) are
        \begin{eqnarray}
        &&|2S,i\rangle =
        \frac{1}{\sqrt 2}(c^\dag_{\alpha i\uparrow}c^\dag_{\beta i\downarrow}
        -c^\dag_{\alpha i\downarrow}c^\dag_{\beta i\uparrow})|0\rangle
         \nonumber \\
        &&|2T1,i\rangle= c^\dag_{\alpha i,\uparrow}
        c^\dag_{\beta i,\uparrow}|0\rangle \label{set}\\
        &&|2T0,i\rangle= \frac{1}{\sqrt
        2}(c^\dag_{\alpha i\uparrow}c^\dag_{\beta i\downarrow}
        +c^\dag_{\alpha i\downarrow}c^\dag_{\beta i\uparrow})|0\rangle
        \nonumber \\
        &&|2T\bar1,i\rangle =c^\dag_{\alpha i,\downarrow} c^\dag_{\beta i,\downarrow}|0\rangle
        \nonumber
        \end{eqnarray}
Here the quantum numbers of two-electron states are qualified as a
singlet $2S$ and triplet $2T\lambda$ with $\lambda=1,0,\bar1$.

The energies corresponding to the eigenstates (\ref{set}) are
defined as follows
\begin{eqnarray}\label{sitr}
E_{1,li}=\epsilon_0,\;\;\; E_{2S,i}=2\epsilon_0 -j,\;\;\; E_{2Ti}=
2\epsilon_0 - I_\perp ,
\end{eqnarray}
where $|\epsilon_0|$ is a single-electron ionization energy,
$j=2t^2_\perp/U$ is the indirect exchange of spins at adjacent
sites, which stems from virtual transitions including double
occupancy. Strong Hubbard repulsion effectively excludes all states
beyond the four states (\ref{set}) in the low-energy part of the
excitation spectrum with $N_i\geq 2$ (see below).

In a charge sector $N_i=1$ the electron spectrum  demands further
diagonalization. The states on a given rung are characterized by
spin  and pseudospin. Spin ${\bf s}_{i}$ is defined as usual, and
pseudospin ${\bf \tau}_i$ is introduced as
\begin{equation}
\tau_{i}^{z}=\frac{1}{2} \sum_\sigma (n_{\alpha i\sigma}-n_{\beta
i\sigma}),~~~ \tau_{i}^{x}=\frac{1}{2}\sum_\sigma \left[
c^\dag_{\beta i\sigma}c_{\alpha i\sigma}+H.c\right]
\end{equation}
The kinetic energy term in (\ref{Hr}) corresponding to the hopping
along the rung  may be cast in the form
\begin{equation}
{\cal K}_\perp=-2t_\perp\sum_i\tau_i^x \label{perpa}
\end{equation}
while, assuming two different electro-chemical potentials
$\mu_\alpha$ and $\mu_\beta$ (or, equivalently, two different
surroundings for $\alpha, \beta$ sites), we get one more term
\begin{equation}
{\cal K}_\mu
=-\sum_i\tau_i^z(\mu_\alpha-\mu_\beta)-\frac{1}{2}\sum_iN_i(\mu_\alpha+\mu_\beta)
\label{mumu}
\end{equation}
in ${\cal H}_r$. Here $N_i=n_{i\alpha}+n_{i\beta}$. Terms in
(\ref{perpa}),(\ref{mumu}) proportional to $\tau^x$ and $\tau^z$
correspond to external uniform "magnetic" field, causing the Zeeman
splitting of the rung orbital states. We discuss the formation of
the orbital gap in the Section IV, taking everywhere below
$\mu_\alpha=\mu_\beta$.

If the sites $\alpha$ and $\beta$ are equivalent, the non-diagonal
transitions described by $\tau_{i}^{x}$ are eliminated in the
local basis of the bonding $(+)$ and anti-bonding $(-)$ states
introduced in accordance with
\begin{equation}\label{humu}
|\pm,\sigma,i\rangle=\frac{1}{\sqrt{2}}(|1,\alpha
i\sigma\rangle\pm|1,\beta i\sigma\rangle).
\end{equation}
At finite $t_\perp$ this basis still diagonalizes the rung
Hamiltonian and the corresponding single-electron energy levels are
$\epsilon_\pm = \epsilon_0 \mp t_\perp $. One should stress that the
doubly occupied states cannot be viewed as two bonding and/or
antibonding orbitals on a rung, due to strong Hubbard repulsion, and
the charge sector $N_i=2$ is represented by the states (\ref{set}).
This is a specific property of CL in comparison with the two-orbital
HC.

Another remarkable property of CL which, in fact, allows double
occupation of a rung without paying large energy $U$, is its
instability against chain dimerization under certain conditions. To
describe this instability, one should introduce the elementary cell
for a CL with doubled lattice period. Such cell is occupied by two
electrons, and its energy spectrum at zero $t_\|$ is easily found.
The low-energy subset is formed by the dimerized states and the
states with singly occupied rungs, respectively:
\begin{eqnarray} \{N_i=2, N_{i\pm1}
=0\}&=&\{|2T\lambda,i; 0,i+1\rangle,|2S,i; 0,i+1\rangle\}\nonumber\\
\{N_i=1, N_{i+1} =1\}&=&\{|\pm,\sigma,i; \pm,\sigma',i+1\rangle\}.
\end{eqnarray}
The corresponding two-electron energy levels are
\begin{eqnarray}
&&E_T(2,0)=2\epsilon_0 - I_\perp , \nonumber \\
&&E_S(2,0)=2\epsilon_0-j, \nonumber \\
&&E_{++}(1,1) = 2(\epsilon_0-t_\perp)\nonumber\\
&&E_{+-}(1,1)= 2\epsilon_0\nonumber\\
&&E_{--}(1,1)=2(\epsilon_0+t_\perp)\label{2set}
\end{eqnarray}
 One concludes from
(\ref{2set}) that the lowest energy levels which predetermine the
spectrum of quarter-filled CL are $E_T(2,0),~ E_S(2,0),~
E_{++}(1,1),~ E_{+-}(1,1).$ If $I_\perp < 2t_\perp$, the lowest
two-electron states of this elementary cell belong to the subset
$\mathbb{S}_{h}=\{+,\sigma,i; +,\sigma', i+1\}$ with one bonding
orbital per rung. In the opposite case $I_\perp
> 2t_\perp$ the lowest  state
correspond to (2,0) occupation of adjacent rungs. We denote the
corresponding subset as $\mathbb{S}_{d}=\{2T,2S,i; 0, i+1\}$. At
finite longitudinal hopping CL forms homogeneous quarter-filled
chain in the former case and dimerized chain with alternating
(2,0) occupation in the latter case.
\begin{figure}
\includegraphics[width=0.25\textwidth]{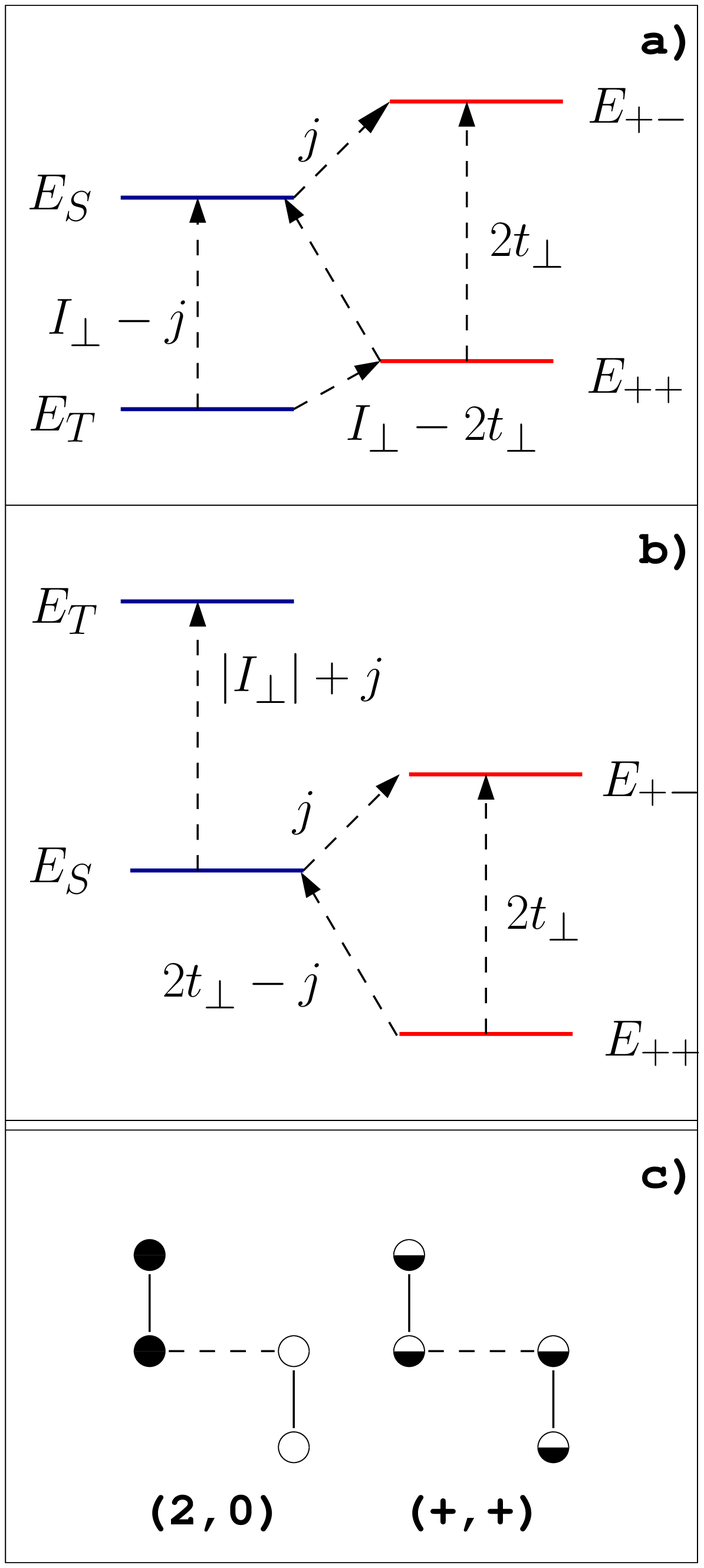}\\
\caption{(Color online) \label{fig:f2a} Two electron level structure
for two neighboring rungs. a) Ferromagnetic on-rung exchange. The
ground state is dimerized. b) Antiferromagnetic on-rung exchange.
The ground state is array of bonding orbitals. c) Occupation scheme
for two different ground states: filled, empty and half-filled
circles correspond to occupied (triplet), empty (vacancy) and
bonding electron states respectively. Transitions within a given
charge spin or charge sector and transitions between different
sectors are marked by vertical and slanting dashed arrows,
respectively. The energies of this transitions are also
shown.}\label{fig:f1a}
\end{figure}

In case of negative (antiferromagnetic) exchange $I_\perp$ the
ground state is homogeneous quarter-filled chain belonging to the
subset $\mathbb{S}_{h}=\{+,\sigma,i; +,\sigma', i+1\}$ and the
lowest excited state is $E_S$ (see Fig. \ref{fig:f1a}b).

In two following sections we discuss charge and spin excitation
spectra of CL  in the limit of weak longitudinal hopping. Looking
at Fig. \ref{fig:f1a}a,b and Eqs. (\ref{2set}), we see that in
case $(a)$ one may find a regime, where only the levels $E_{++}$
and $E_T$ are involved in formation of the low-energy excitations,
whereas in case $(b)$ at least three states $E_{++}, E_S, E_{+-}$
are involved.

\section{Weak longitudinal hopping limit, Ferromagnetic on-rung exchange}
In this section we analyze the quarter-filled CL with $I_\perp>0$
in a situation , where the longitudinal hopping is small in
comparison with the on-rung coupling constants, and the
inequalities
\begin{equation}\label{ineq1}
I_\perp > 2t_\perp, ~~~ t_\| \ll I_\perp- 2t_\perp .
\end{equation}
are valid. In the absence of longitudinal hopping each second rung
is empty and all doubly occupied rungs are in spin one triplet
states (see Fig. \ref{fig:f1a}a). One has to study longitudinal
spin and charge excitations at finite $t_\|$ above the dimerized
ground state with alternating rungs $\ldots |2T\lambda,i\rangle,
|0,i+1\rangle \ldots$ The charge excitations arise due to
transitions $\mathbb{S}_{d} \Leftrightarrow \mathbb{S}_{h}$. These
excitations (charge-transfer excitons) are obviously gapped, so
that the low-energy excitations belong to the spin sector.

In zeroth order approximation the spectrum of rung dimers is
determined by the eigenstates of the Hamiltonian (\ref{Hr}) in a
charge sector $N_i=2$ for FM on-rung exchange. These dimers are in
the state of {\it spin rotator},\cite{kak05a,kak05b,KA01}
characterized by the singlet-triplet manifold
$\{|2T\lambda,i\rangle,|2S,i\rangle\},$ with the energy gap between
two spin states given by
\begin{equation}
\Delta E_{ST} =I_\perp- j  .
\end{equation}
Longitudinal hopping (\ref{Hl}) generates indirect exchange
produced by the electron cotunneling via empty odd rungs.
 To derive the corresponding  spin Hamiltonian, we have to exclude the
charge sector $\mathbb{S}_{h}$ from the effective phase space. It is
seen from Fig. \ref{fig:f1a}a, that the singlet states are
effectively quenched at the energy scale $I_\perp-2t_\perp$, which
characterizes the charge transfer gap, so that one may confine
oneself with the subspace $\{{\mathbb S}^\prime_d, {\mathbb S}_h
\},$ with singlet states excluded from the subset ${\mathbb
S}^\prime_d=\{2T,i; 0, i+1\}$ of "polar" states.

\begin{figure}
\includegraphics[width=0.17\textwidth]{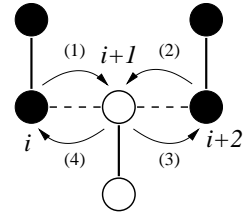}\\
\caption{\label{fig:f2b} Scheme of superexchange induced in 4-th
order hopping processes in dimerized CL. Arrows indicate sequential
electron hops resulting in the intermediate states (\ref{virt}).
Notations for the occupation scheme are the same as for Fig.
2c.}\label{fig:f1b}
\end{figure}

The remarkable property of the state ${\mathbb S}^\prime_d$ is
that the macroscopic configurational degeneracy of the Hubbard
model with partial filling is removed due to dimerization. The
only remaining even-odd degeneracy of the dimerized ground state
of zero-order Hamiltonian may be also removed in case of CL with
odd number of rungs $M=2N+1$, where, say, all even rungs
$0,2,\ldots 2N$ are doubly occupied and all odd rungs $1,3\ldots
2N-1$ are empty. Then only the trivial spin degeneracy remains,
and one may strictly define the effective exchange Hamiltonian in
the 4-th order of Brillouin-Wigner expansion
\begin{equation}
{\cal H}_{eff} = {\cal H}_r+ \langle \Psi_0|{\cal
H}_l^{(4)}|\Psi_0\rangle,
\end{equation}
where the brackets $\langle \Psi_0|\ldots|\Psi_0\rangle$ stand for
the dimerized polar ground  state with alternating triplet spin
states and vacancy states
\begin{equation}\label{h4}
{\cal H}_l^{(4)}= \frac{{\cal
H}_{l}|\phi^{(1)}\rangle\langle\phi^{(1)}|{\cal
H}_{l}|\phi^{(2)}\rangle\langle\phi^{(2)}|{\cal
H}_{l}|\phi^{(3)}\rangle\langle\phi^{(3)}|{\cal H}_{l}}
    {(-E_0^{(1)})(-E_0^{(2)})(-E_0^{(3)})}
\end{equation}
Here the projection operators
$$
P^{(\alpha)}=|\phi^{(\alpha)}\rangle\langle\phi^{(\alpha)}|,~~~
\alpha=1,2,3
$$
include the states which are generated from the ground state
$|\Psi_0\rangle$ after 1-st, 2-nd and 3-rd action of the operator
${\cal H}_{l}$, respectively and $E_0^{(\alpha)}$ are the
corresponding excitation energies. The 4-th order cotunneling
process resulting in effective kinematic exchange is illustrated by
Fig. \ref{fig:f1b}. It is evident that only the clusters $(i, i+1,
i\pm 2)$ are involved in formation of the effective exchange
Hamiltonian where $i$ stands for any even site. We consider here the
simplest situation, where the singlet intermediate states can be
discarded in the main order, which happens at $t_\|/(I_\perp-j) \ll
t_\|/(I_\perp -2t_\perp)$. This simplification does not change our
qualitative conclusions.
\begin{eqnarray}\label{virt}
    |\phi^{(1)}\rangle &=& |+\sigma_1,i;~+\sigma_2,i+1;~
2T\lambda,i+2\rangle \nonumber \\
    |\phi^{(2)}\rangle &=& |+\sigma_1,i;~2T\lambda_1,i+1;~
+\sigma_3,i+2\rangle \\
    |\phi^{(3)}\rangle &=& |+\sigma_1,i;~+\sigma_4,i+1;~
2T\lambda',i+2\rangle \nonumber
\end{eqnarray}

It is seen from Eq. (\ref{virt}) and Fig. \ref{fig:f1b}, that the
intermediate states involve electron cotunneling exchange between
two nearest even sites with formation of a virtual doubly occupied
triplet state on the odd site between them. This mechanism is
known as superexchange in the theory of magnetic dielectrics.
Appearance of intermediate triplet states $|2T\lambda\rangle$ in
the indirect exchange interaction is the peculiarity of CL: unlike
the two-orbital HC model, double occupation of intermediate state
does not cost the Hubbard repulsion energy $U$, because the two
electrons on a rung may occupy different sites.
As a result we come to the following effective spin Hamiltonian
\begin{equation}
{\cal H}_{eff}={\cal H}_{r}-J_\|\sum_i {\bf S}_i \cdot {\bf S}_{i+2}
\label{FMeffmodel}
\end{equation}
where the site index $i$ stands only for even rungs of CL. The
coupling constant is
\begin{equation}\label{ext}
J_\| = \frac{t^4_\|}{16\Delta_T^3}
\end{equation}
where $\Delta_T=I_{\perp}-2t_\perp$ is the charge transfer energy
gap (see Fig. \ref{fig:f1a}a). The details of calculation are
provided in Appendix. Deriving (\ref{ext}) we assumed that the
singlet state does not contribute to superexchange. This assumption
holds provided $\Delta_T\ll \Delta E_{ST}$.

Thus we have demonstrated that the problem of quarter-filled CL is
mapped onto the familiar problem of spin-one Heisenberg chain with
ferromagnetic exchange in the limit of weak longitudinal hopping
under conditions (\ref{ineq1}). Only the even sites are involved
in formation of the gapless spin excitations, which, apparently
may be described in terms of the spin-wave theory.\cite{ivan}

As to the charge transfer excitons, $|+\sigma,i;
+\sigma',i+1\rangle$, shown in the right column of Fig.
\ref{fig:f1a}c, these states, like other excitons can exist both
in triplet and singlet spin configurations. These excitations may
propagate coherently by means of cotunneling reminding that
described by Eqs. (\ref{h4}), (\ref{virt}). To provide translation
of such exciton from one cell of dimerized CL to another, two
electrons should hop to the right $(i\to i+1; i+1 \to i+2)$ and
then to the left $(i+2\to i+1; i+3 \to i+2)$. This 4-th order
indirect "exchange" provides excitons with the dispersion law
$\omega_q\sim \Delta_T+ 2J_\|\cos 2qa$, where $a$ is the lattice
spacing along the leg.

Thus, we have shown in this section that the spin excitations in the
CL with ferromagnetic on-rung exchange are gapless magnon-like
modes, whereas the  excitations in a charge-sector are gapwise
excitons describing coherent propagation of two bonding states along
the chain.

\section{Weak longitudinal hopping limit, Antiferromagnetic on-rung exchange}

\begin{figure}
\includegraphics[width=0.35\textwidth]{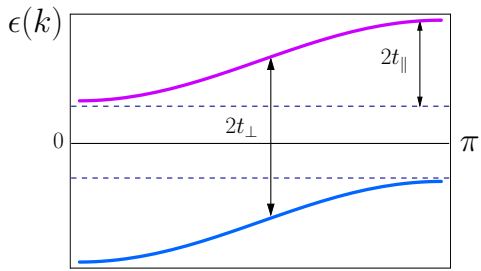}\\
\caption{(Color online) \label{fig:f3} Tight-binding band structure
for quarter filled centipede model at $t_\|\ll t_\perp$. The
occupied and empty orbital  subbands $\epsilon_{\pm}(k)$ are shown.
The chemical potential is chosen as a reference energy level.}
\end{figure}

In case of negative on-rung exchange hopping along the leg and
small $t_\|$, so that the inequality
\begin{equation}\label{ineq2}
t_\| \ll 2t_\perp -j
\end{equation}
is satisfied, the energy levels are ordered in accordance with
Fig. \ref{fig:f1a}b, and the ground state wave function of
quarter-filled CL is given by the product of bonding orbitals
\begin{equation}
 |\Psi_0\rangle  =| \ldots
+ \sigma_i,i;~ +\sigma_{i+1},i+1;~ +\sigma_{i+2},i+2; \ldots
\rangle .
\end{equation}
In this sector of  phase diagram, the quarter-filled CL is close
in a sense to the quarter-filled two-orbital HL,\cite{Yama} with
lifted orbital degeneracy: each rung is occupied by one electron
with bonding orbital, and the empty antibonding level is separated
from this band by the orbital gap $\sim 2t_\perp$ [see Eqs.
(\ref{perpa}), (\ref{mumu}) and Fig. \ref{fig:f3}].

The important distinction from two-orbital HL roots in the
structure of the doubly occupied states. Since two electrons on a
rung occupy different orbitals in accordance with (\ref{set}), the
energy cost of this state is not the Hubbard repulsion energy $U$
but the difference $\Delta_S=E_S-E_{++}=2t_\perp -j$, which is
even smaller than the orbital gap. Nevertheless, magnetic
properties of these two models are similar in the limiting case
(\ref{ineq2}): the longitudinal hopping generates effective AFM
longitudinal exchange $\sim t_\|^2/\Delta_S$, and the gapless spin
liquid state of the resonance valence bond (RVB) type is realized.

The charge excitations exhibit more interesting character, as we
discuss below. The lowest excited state is a  vacancy-singlet pair
occupying neighboring rungs
\begin{equation}
    |S,i; 0,i+1\rangle= |\ldots +\sigma,i-1;~
 S,i;~0,i+1;~ +\sigma,i+2 \ldots \rangle
\end{equation}
with the energy $\Delta_{S}$.

Clearly, the state $|S,i; 0,i+1\rangle$ is an analog of the charge
transfer exciton briefly discussed in the end of section II.
However, in this case the exciton propagates via intermediate
bonding-antibonding levels $\epsilon_{\pm}$ due to the strong
inequality $t_\perp \gg j$.
\begin{figure}
\includegraphics[width=0.15\textwidth]{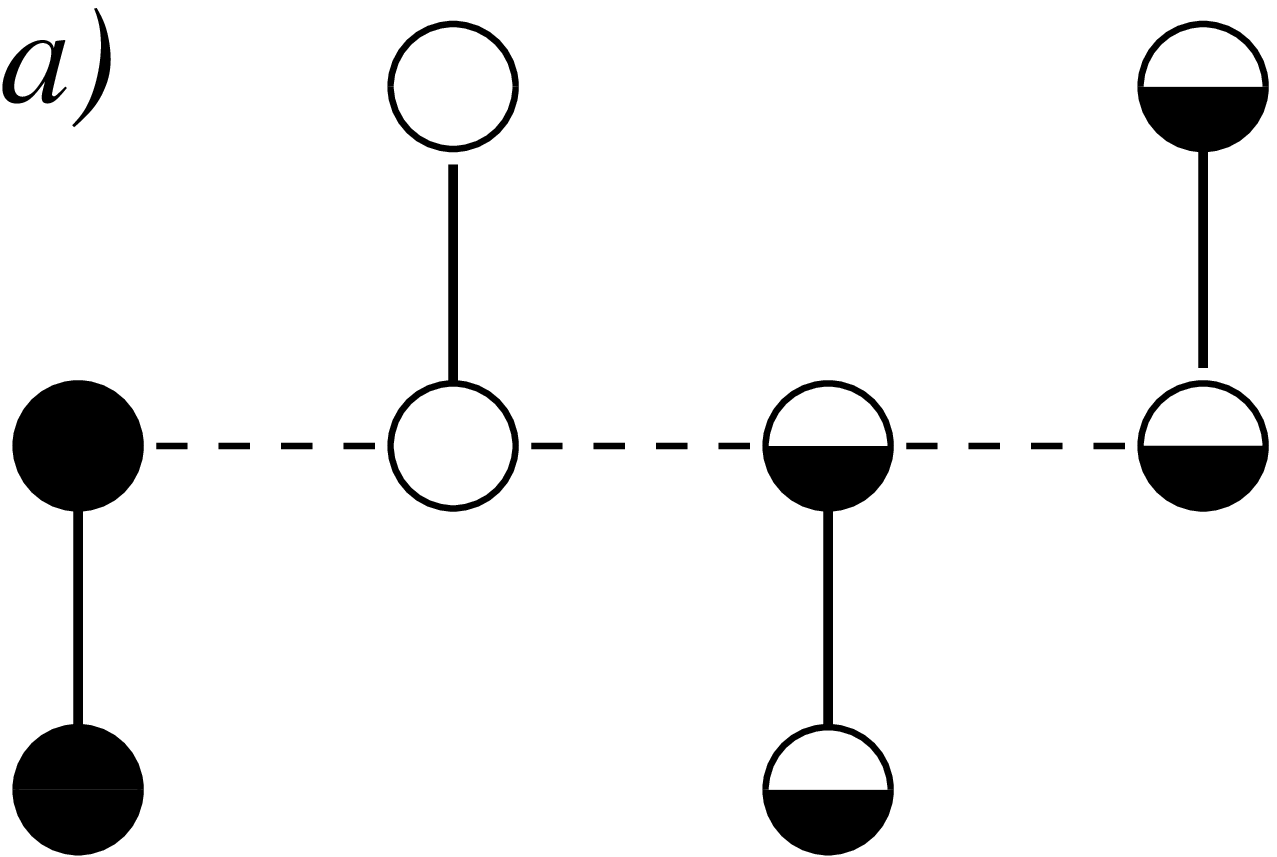}\hspace*{8mm}%
\includegraphics[width=0.15\textwidth]{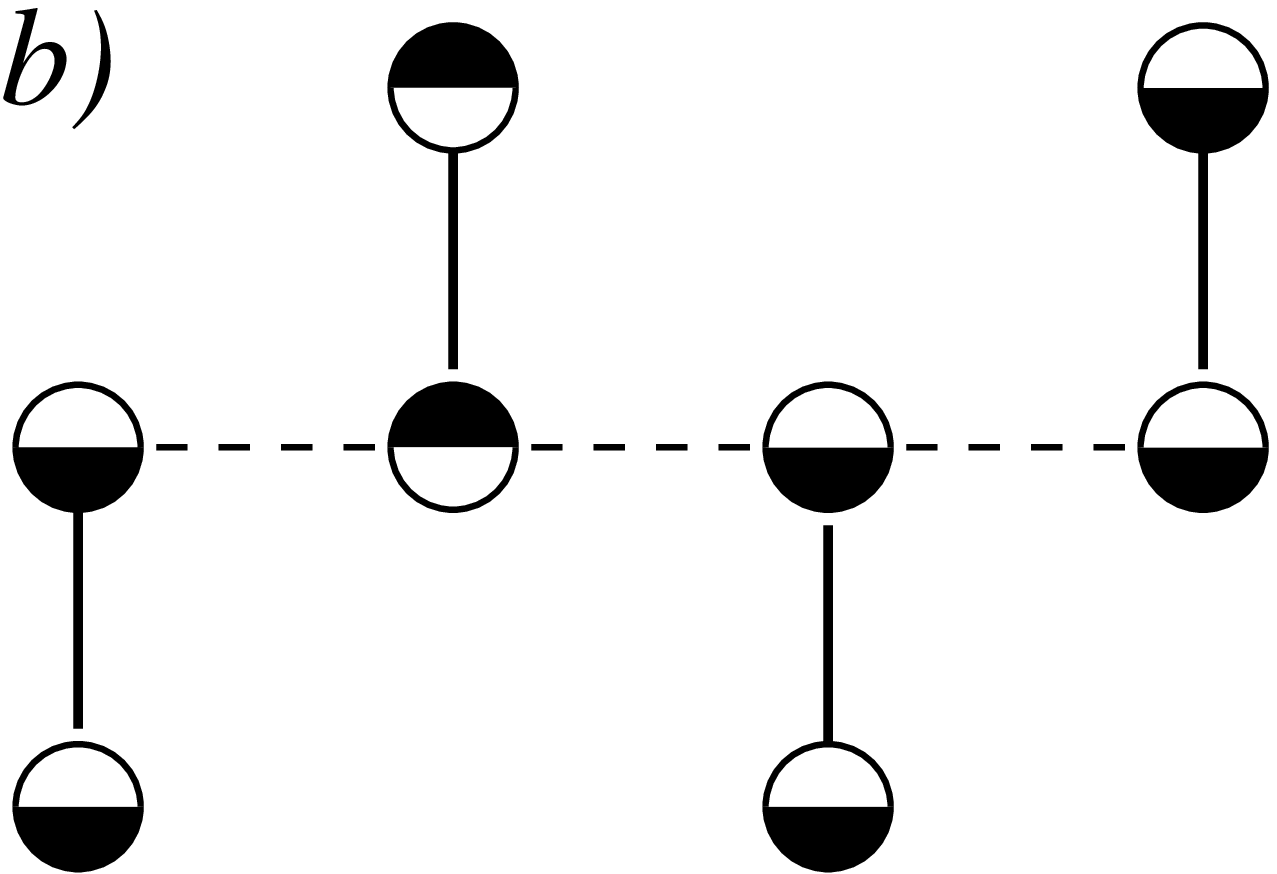}\\\vspace*{8mm}
\includegraphics[width=0.15\textwidth]{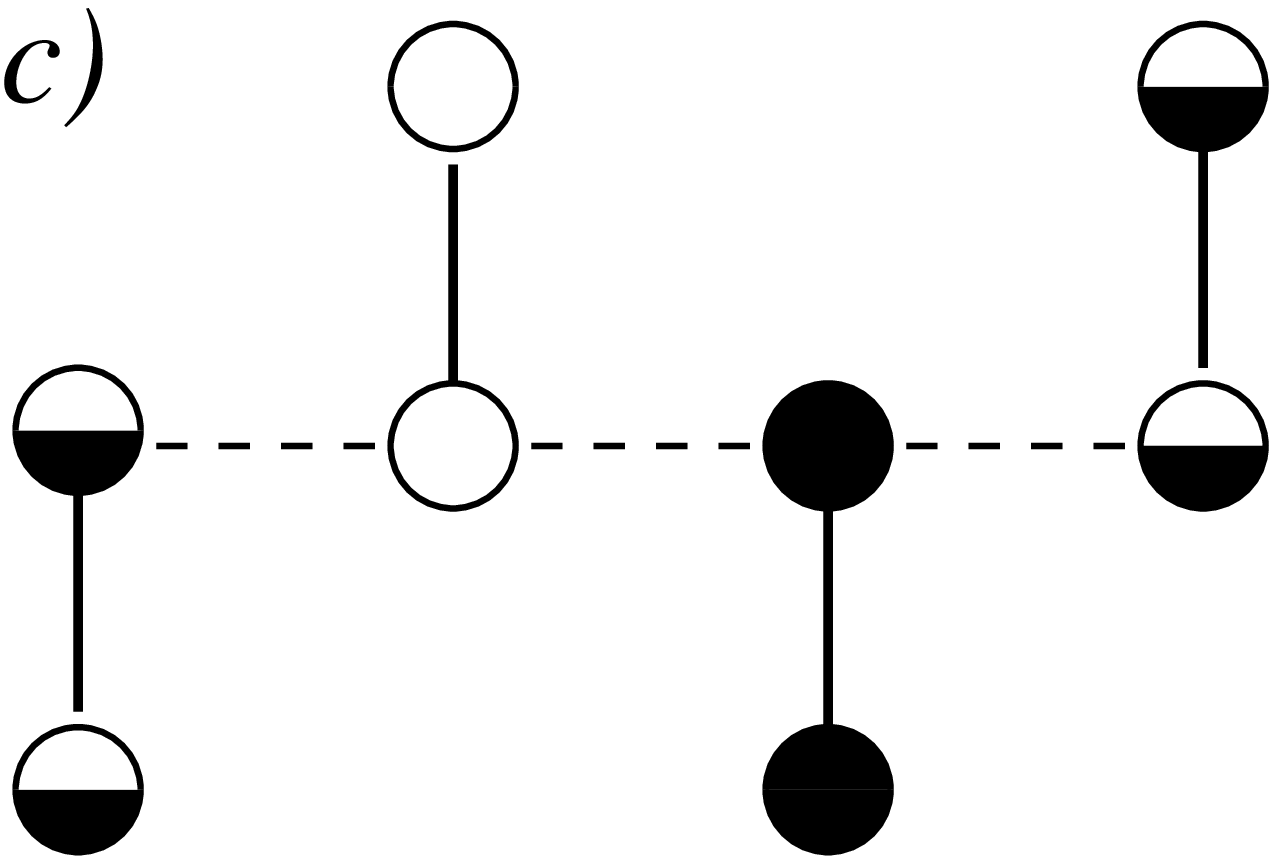}\hspace*{8mm}%
\includegraphics[width=0.15\textwidth]{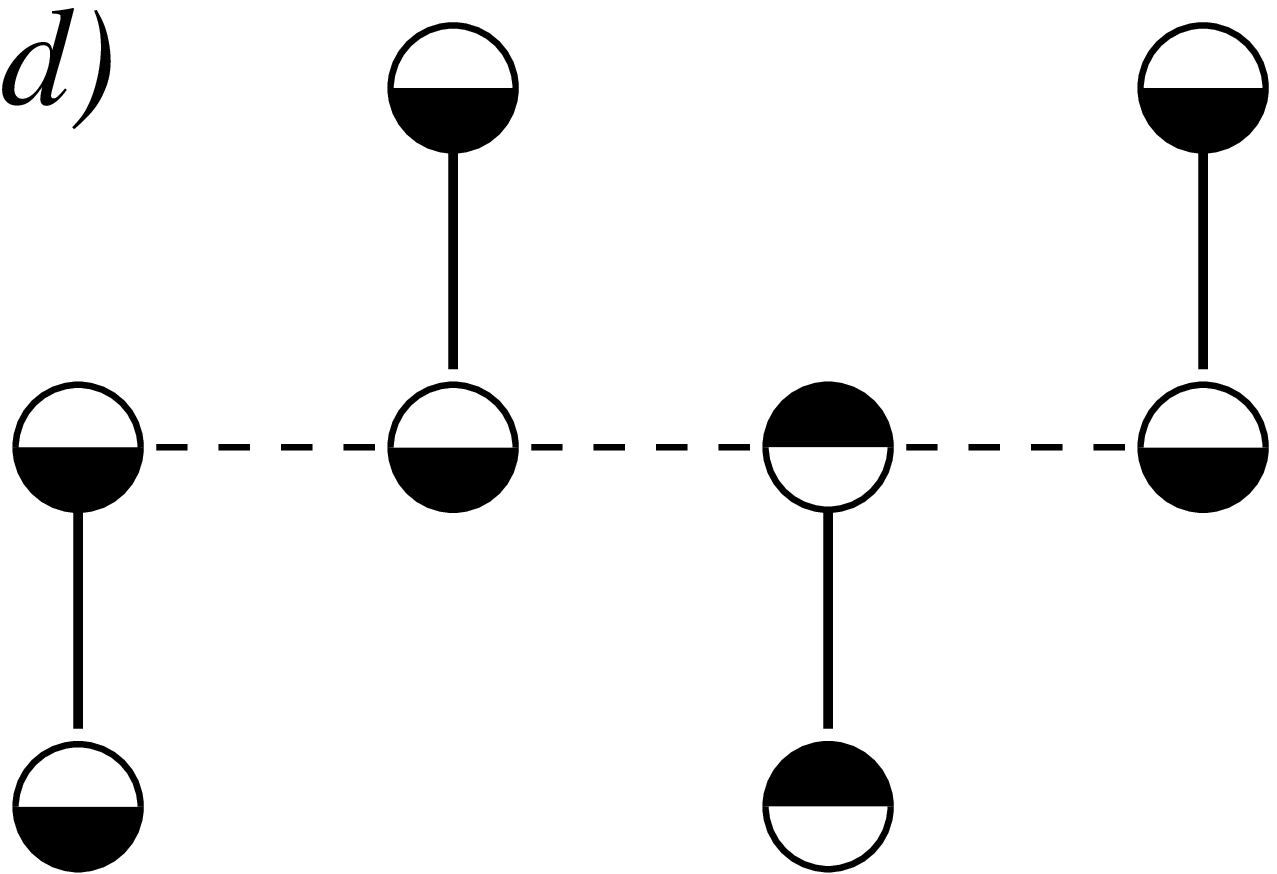}\\\vspace*{8mm}
\includegraphics[width=0.15\textwidth]{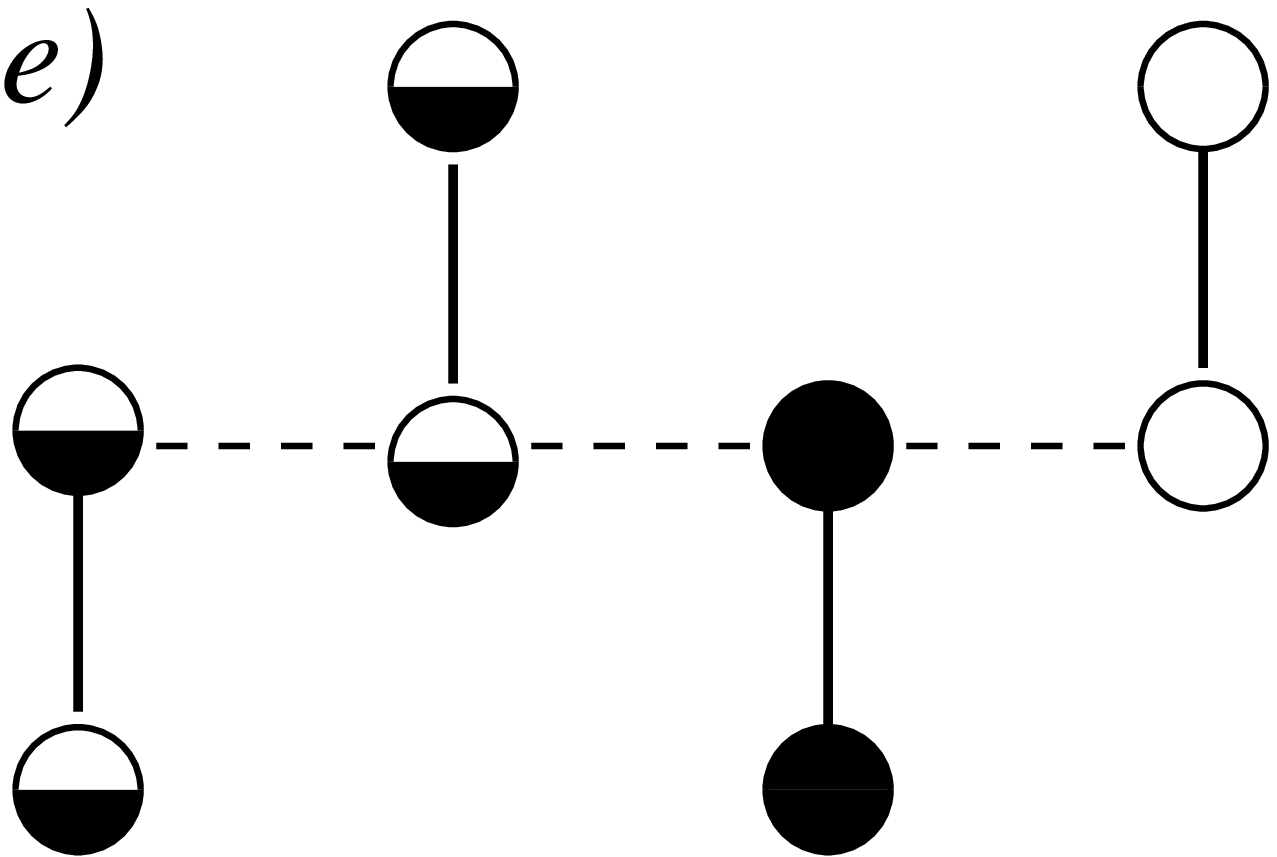}
\caption{\label{fig:f3a} Coherent motion of the vacancy-singlet
exciton due to the 4-th order electron hopping $(1\to 2;~2\to
3;~3\to 2;~4\to 3)$. Diagrams  illustrate electron configurations in
the initial ${\sl (a)}$, subsequent intermediate ${\sl (b)-(d)}$ and
finite ${\sl (e)}$ states of exciton translation. Filled, empty,
half-filled (black down) and half-filled (black up) circles
correspond to singlet, vacancy bonding and antibonding electron
states, respectively.}
\end{figure}
Figure \ref{fig:f3a} illustrates the mechanism of exciton
propagation. Here the rung occupied by the antibonding electron is
shown by the pair of half-filled circles with filled upper half.
Then the elementary hopping from doubly occupied to empty rung may
be presented as
\begin{equation}
|S,i;0,i+1\rangle \to |\pm,i; \mp, i+1\rangle~,
\end{equation}
and it costs the energy $j$, according to (\ref{2set}). Note that a
crucial detail allowing one to discuss exciton propagation as the
charge excitation is the fact, that the spin sequence of the bonding
orbitals in the initial and final state in Fig.\ \ref{fig:f3a} is
unchanged and the spinon dispersion is unaffected by the coherent
motion of the singlet exciton in the long wave limit. This is a
manifestation of the spin charge separation in quarter-filled CL
model.

In the limit of extremely weak longitudinal hopping $t_\| \ll j$,
the singlet-vacancy pair may propagate coherently. This pair is
doubly degenerate due to its permutation symmetry, and this
degeneracy manifests itself in existence of right- and left-moving
charge-transfer excitons. The 4-th order hopping process resulting
in translation of the right-mover between two adjacent double cells
is shown in Fig. \ref{fig:f3a}. Superposition of right- and
left-moving excitons results in the dispersion law
\begin{equation}
\omega_q\sim \Delta_S - J'_\|\cos 2qa\label{movex}
\end{equation}
with $J'_\|\sim t^4_\|/j^3$. We showed before that the spin sequence
remains unchanged (Cf. Ref.\onlinecite{grar}) upon the motion of
exciton, and the gain in the kinetic energy is $\sim J'_\|$. The
resulting spin configuration in Fig.\ \ref{fig:f3a} corresponds to
two spins detached from the main spin sequence on the right and
attached to one on the left. For long-wave modes of exciton motion,
$q\to 0$ in (\ref{movex}), the change in the magnetic energy
associated with this disruption is negligible compared to the gain
$\sim J'_\|$, and charge excitons move coherently.

\begin{figure}
\includegraphics[width=0.15\textwidth]{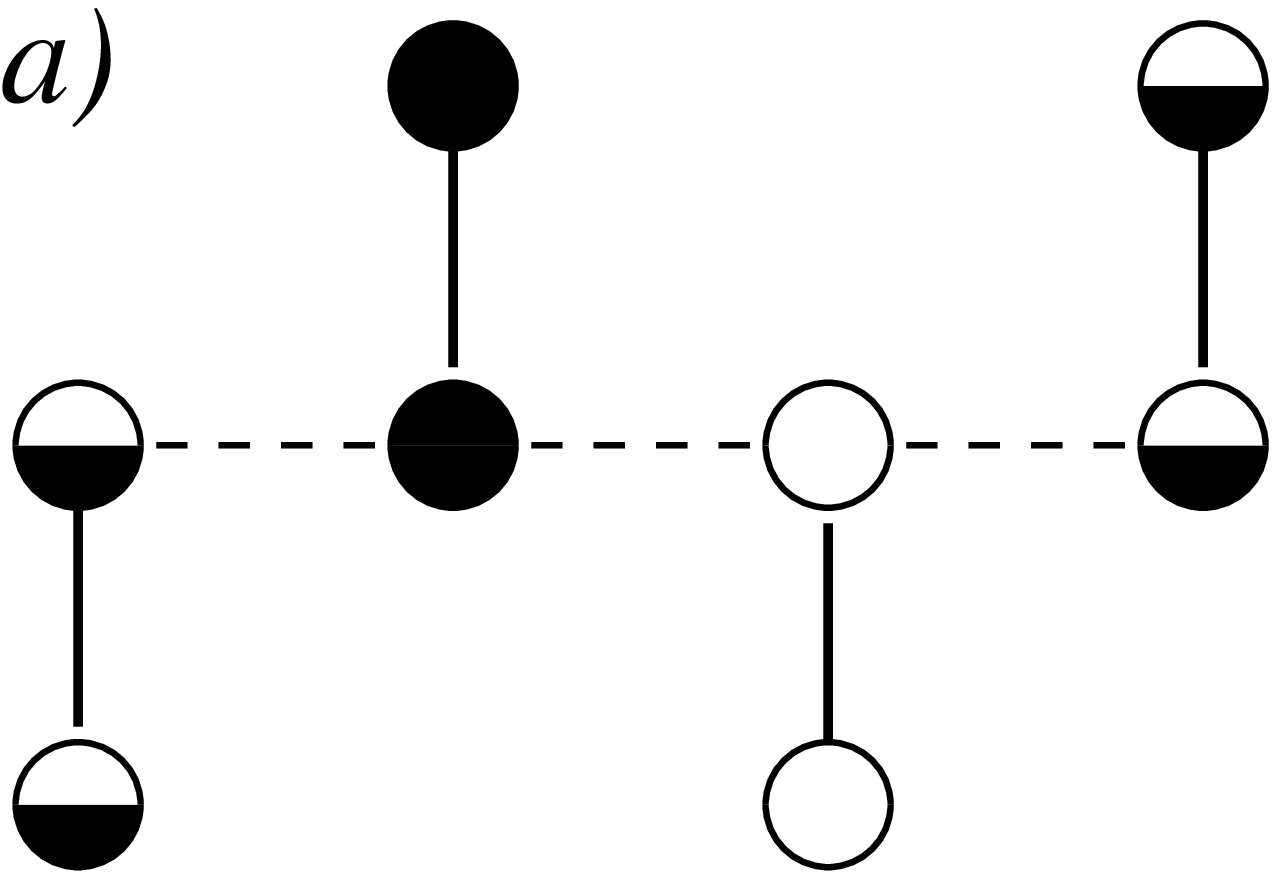}\hspace*{8mm}%
\includegraphics[width=0.15\textwidth]{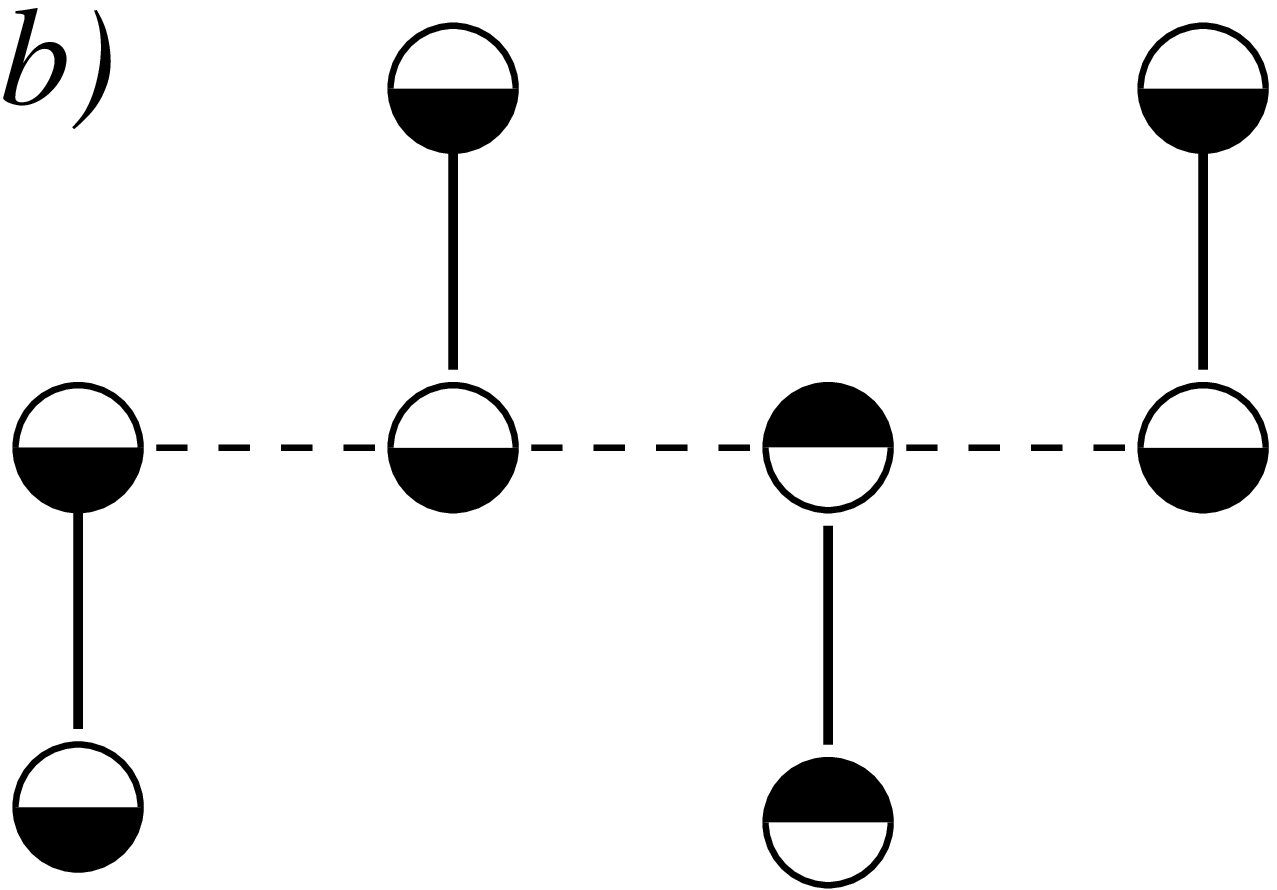}\\\vspace*{8mm}
\includegraphics[width=0.15\textwidth]{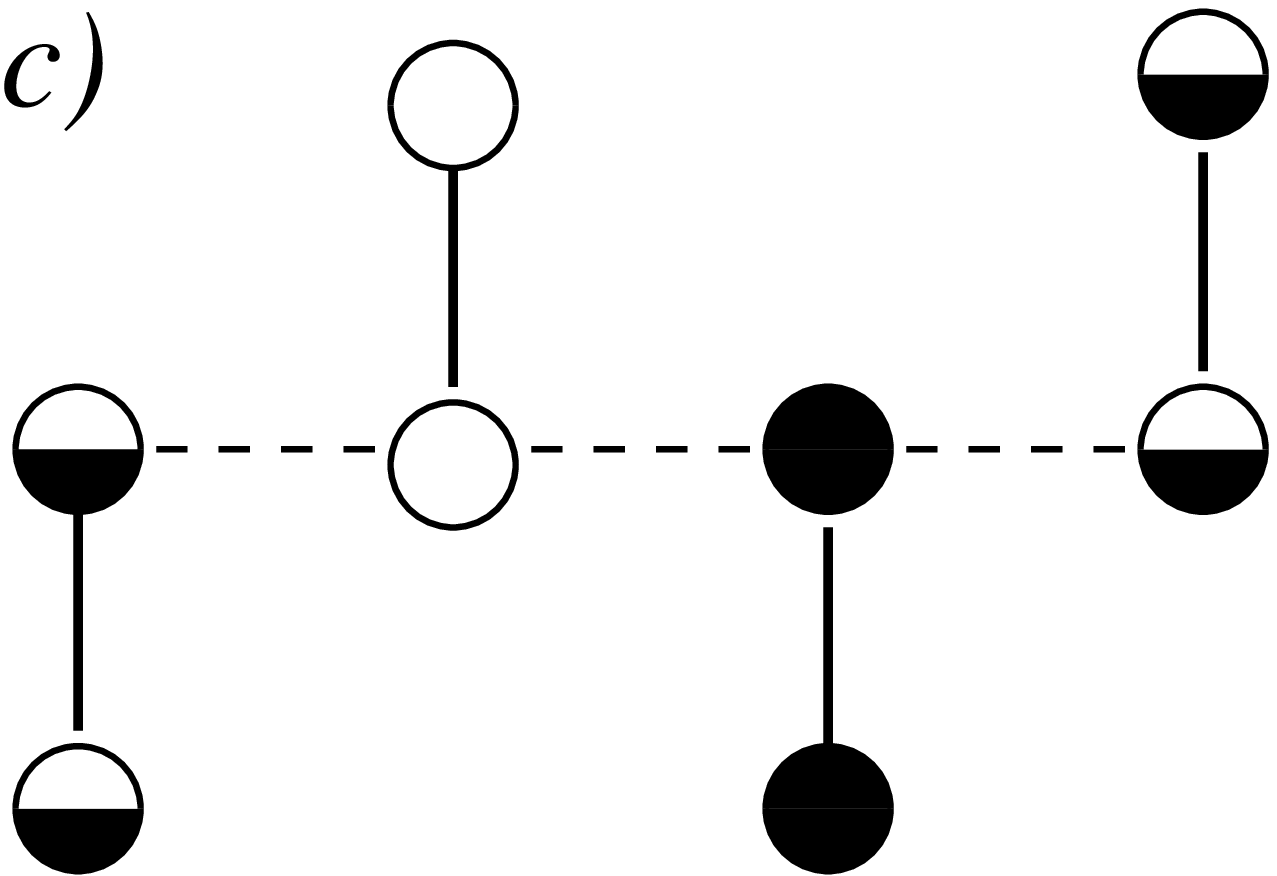}\hspace*{8mm}%
\includegraphics[width=0.15\textwidth]{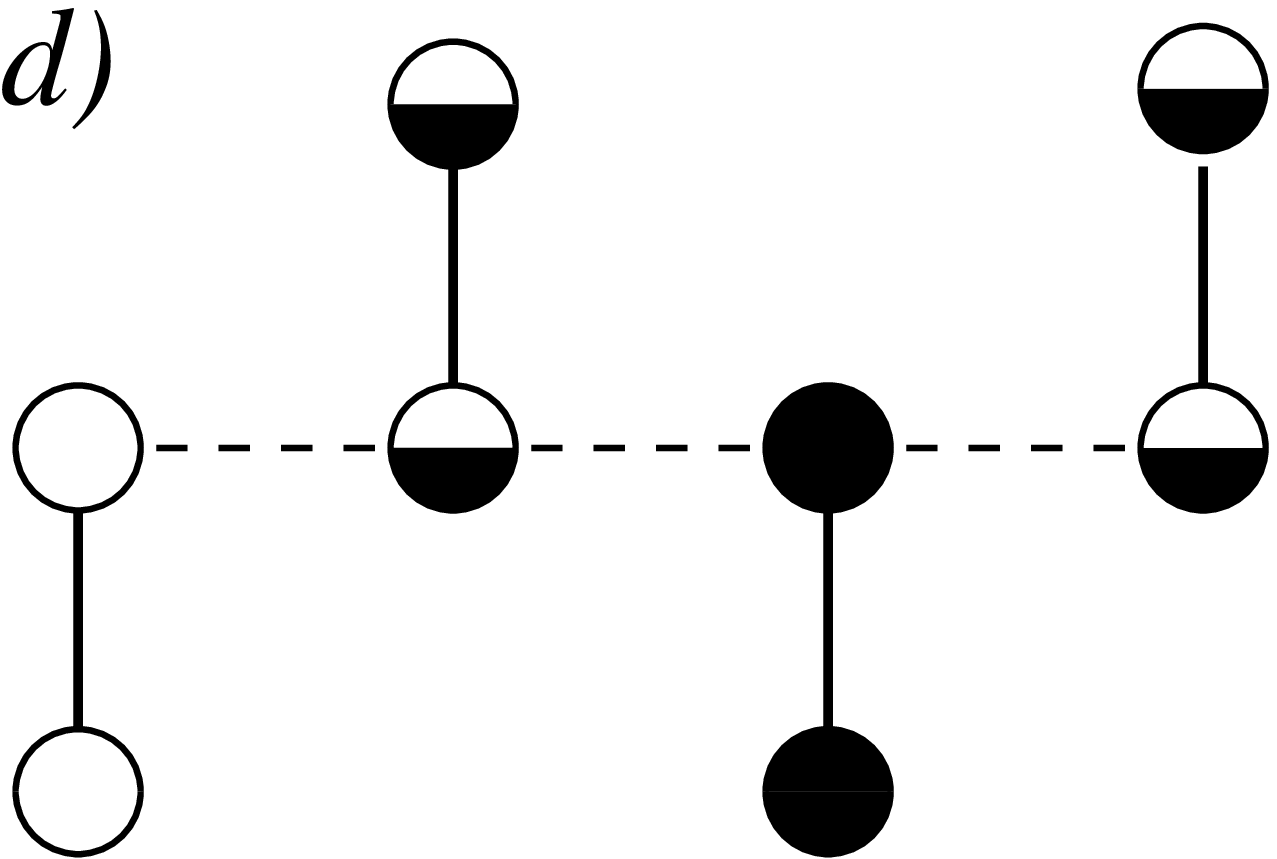}\\\vspace*{8mm}
\includegraphics[width=0.15\textwidth]{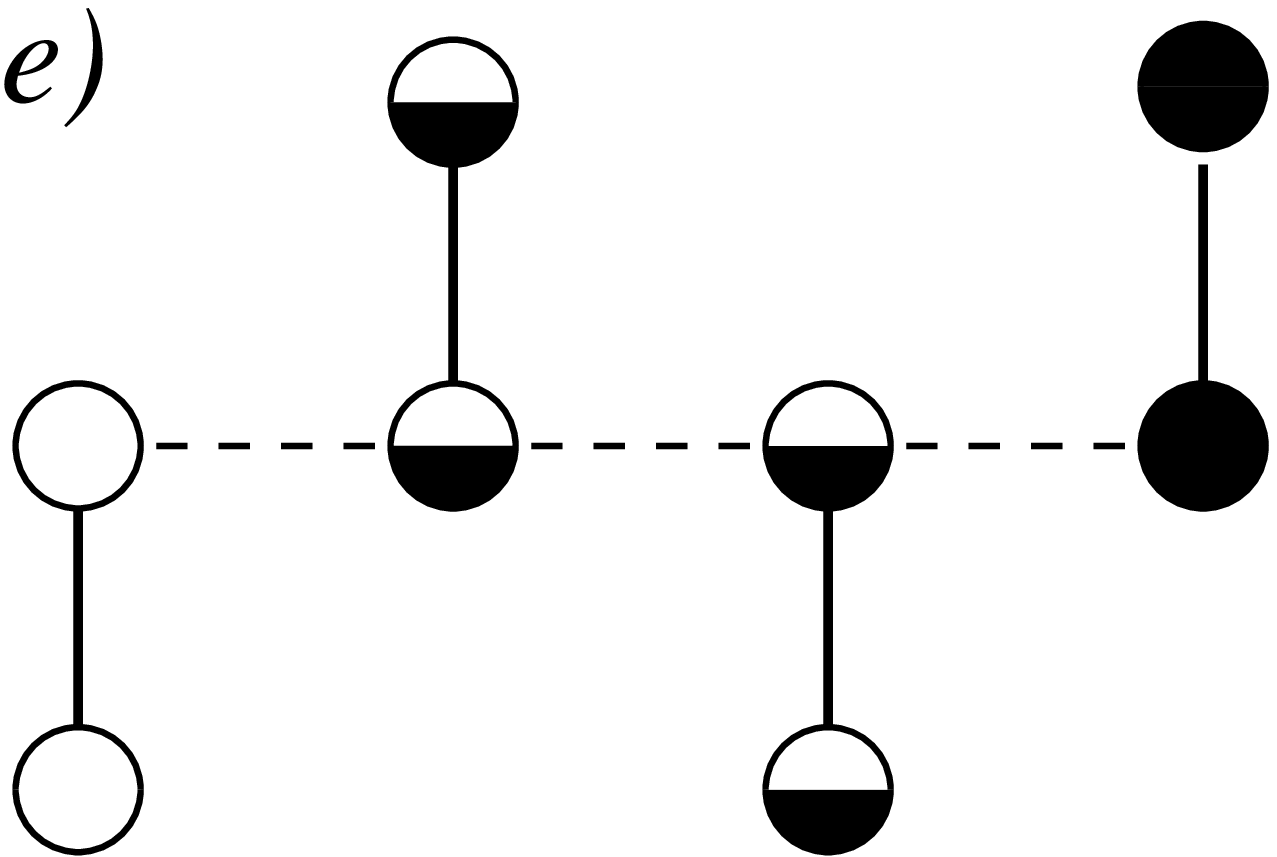}
\caption{\label{fig:f3b} Incoherent  motion of the vacancy and
singlet due to the 4-th order electron hopping $(2\to3; 2\to3;
1\to2; 3\to 4)$. Notations for the occupation scheme are the same as
for Fig.5.}
\end{figure}

Now we turn to the case  $t_\parallel > j$.  In this situation the
coherent  "vacancy-singlet" excitons are unstable against
dissociation into independently propagating singlet and vacancy
"defects" due to the overlap of the bands originating from the
levels $E_S$ and $E_{+-}$. Apparently, only incoherent propagation
of such defects is possible. An example of 4-th order process
resulting in decay of singlet-vacancy pair is presented in Fig.
\ref{fig:f3b}. Such incoherent processes will be accompanied by
spin-flips, antibonding-bonding de-excitation and eventual
relaxation towards the ground state.

However, coherent motion of defects is possible at small deviation
from one-quarter filling.  Excess holes (vacancies) behave similarly
to the holes in the nearly half-filled HC in the limit
(\ref{ineq2}), where the longitudinal hopping does not result in
excitation of the upper orbital subband $\epsilon_-$. In this case
one deals with a fermion gas,``fully polarized''  with respect to
the pseudospin $\tau$. Excess electrons fill the side sites on the
rungs without paying the Hubbard energy $U$, and it is easy to check
that the hopping along the leg is not accompanied by creation of a
spin string because the doubly occupied state is a spin singlet. As
a result, the coherent propagation of these defects is also possible
under the conditions (\ref{ineq2}), when the antibonding orbitals
are not excited. Due to specific geometry of CL, there is no
particle-hole symmetry near the 1/4 filling, and the hopping rates
for particle and hole propagation are different. Direct calculation
shows that the hopping rate  for coherent motion of a vacancy
reduces due to the kinematic constraint from $t_\|$ to $\tilde
t_v=t_\|/2$, whereas the similar reduction for the singlets gives
$\tilde t_s=t_\|/4$. Thus, basically the coherent motion of the
vacancy and singlet can be considered in terms of tight-binding
bands $\epsilon_{1,2}(k)=-2\tilde t_{1,2} \cos k$. Evidently,
singlet (vacancy) cannot occupy the same rung twice, hence the
corresponding excitations are fermion-like. We can think of them as
of spinless fermions characterized by two colors $f_{i,v}$ and
$f_{i,s}$:
\begin{equation} |0,i\rangle\langle +,i| \to f^\dagger_{i, v},~~~
|S,i\rangle\langle +,i| \to f^\dagger_{i, s}
\end{equation}

Even at exact quarter filling but at finite temperatures, the
vacancies and singlets may appear at small concentrations due to
thermal activation. These  excitations are described by the
effective two color Fermi liquid model
\begin{equation} H_{eff} = \sum_{k,\alpha=v,s}\epsilon_\alpha(k) f^\dagger_{k,\alpha}
f_{k,\alpha} +K \sum_j f^\dagger_{j,v} f_{j,v} f^\dagger_{j,s}
f_{j,s}
\end{equation}
where the effective constant $K$ describes effective short-range
repulsion between two fermions, which prevents them from occupying
the same site. This constant may be estimated as $K\sim
t_\|^2/\Delta_S$, which is small comparing to the band-widths
$\sim t_\|$. Having in mind the  asymmetric hopping Hubbard model
\cite{lieb}, it is tempting to analyze it using the
well-established machinery of bosonization etc. However, the
sizeable concentrations of vacancies and singlets are expected at
temperatures comparable to the main parameter, $t_\perp$. This
argument shows that that fermion gas in $H_{eff}$ is not
degenerate, and the linearization of the spectrum around Fermi
points is unjustified.

As to the excitations in the spin sector, one may say that the
problem of spin excitations in presence of a singlet or a vacancy
may be mapped on the problem of RVB spin liquid with spin-one
defect or dangling bond, respectively (see Ref. \onlinecite{Egaf}
for general approach).

To summarize, we have found in this section, that the spin
excitations are gapless  Fermi-like spinons both at exact 1/4
filling and at small deviation from this point. As to the coherent
charge excitations, these  are gapped excitons at extremely small
$t_\|\ll j$ and gapless fermions at finite doping and/or finite
temperature.

\section{Strong longitudinal hopping}

The CL model possesses unusual properties also in the limit of
strong longitudinal hopping, when it dominates over the
transversal degrees of freedom:
\begin{equation}
t_\|\gg   t_\perp \gg I_\perp,
\end{equation}
In this case the on-rung excitations shown in Fig. \ref{fig:f1a}
are completely smeared, and the structure of low-energy spectrum
of the system is predetermined by orbital degrees of freedom
(bonding-antibonding subbands) and corresponding Hubbard
repulsion. We will show in this section that in this simple
situation close to the two-orbital HC the charge excitations
behave unconventionally because of the gap opening and
corresponding van Hove singularities appearing in the lower
partially occupied Hubbard band.

For a clear presentation we compare the case of the asymmetric
two-leg (2L) ladder and centipede ladder (see Fig.
\ref{fig:f1}b,a, respectively. Fourier transforming along the leg,
we write the kinetic part of the Hamiltonian as
        \begin{eqnarray}
        H_{kin} &=& -\Psi_{k}^\dagger
        \begin{bmatrix}
        2 t_\| \cos k +  \mu, &   t_\perp \\
        t_\perp , &   2P t_\| \cos k -\mu
        \end{bmatrix}
        \Psi_{k}       \\       \Psi_{k}^\dagger &=& (
        \tilde c_{\alpha,k,\sigma}^\dagger,
        \tilde c_{\beta,k,\sigma}^\dagger )
        \end{eqnarray}
where $P =1$ for the 2L model and  $P =0 $ for CL model. At this
point we also allow for the difference in the electrochemical
potentials, $2\mu=\mu_\alpha-\mu_\beta$, between the sites
$\alpha$ and $\beta$ [see Eq. (\ref{mumu})]. The dispersion in
these two extreme cases is given by

        \begin{eqnarray}
        &&\varepsilon_\pm(k) =
        - 2 t_\| \cos k \mp \sqrt{\mu^2 +t_\perp^2 }  ,
        \quad  \mbox{2L} \label{sqr}
        \\ &=&
        - t_\| \cos k \mp
        \sqrt{t_\perp^2 + (\mu + t_\| \cos k)^2},
        \quad \mbox{CL}
        \label{disp-oneleg}
        \end{eqnarray}

\begin{figure}
\includegraphics[width=0.4\textwidth]{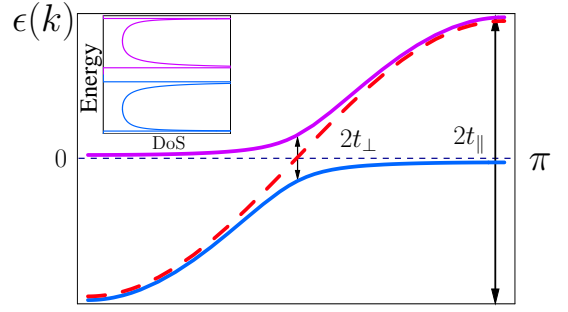}\\
\caption{(Color online) \label{fig:f4} Tight-binding band structure
of two lowest Hubbard subbands for quarter filled centipede model at
$t_\perp\ll t_\|$. The electro-chemical potential $\mu_a=\mu_b$. The
direct gap corresponds old Fermi surface point $k_F=\pi/2$. The
indirect gap $\Delta_{ind}$ is between the bottom of conduction band
and the top of the valence band is characterized by the "flat" parts
of the spectra. The inset shows the energy dependence of the Density
of States (DoS). The character of the DoS singularities and its
influence on the response functions is discussed in the Section IV.}
\end{figure}
The square root term in Eq. \ref{sqr} has a meaning of effective
Zeeman splitting. In contrast to 2L situation where the bands
$\epsilon_+(k)$ and $\epsilon_-(k)$ bands overlap we see that in
CL case two branches of dispersion are always separated by the gap
(Fig. \ref{fig:f4}). This indirect gap is given by
\begin{equation}\Delta_{ind} =2(
\sqrt{t_\perp^2 + t_\|^2} -t_\|) \simeq t_\perp^2
/t_\|.
\end{equation}
for the case of equal electrochemical potentials ($\mu=0$) or
\begin{equation} \Delta_{ind} \simeq
t_\perp^2 t_\|/(t_\|^2 - \mu^2)
\end{equation}
for $\mu_\alpha\neq\mu_\beta$. The direct gap is realized at
$k=\pi/2$ for $\mu =0$ and at $k = \arccos(-\mu/t_\|)$ otherwise
and is equal to $\Delta = 2 t_\perp$ in both cases.  For exact 1/4
filling, the overall chemical potential
$\mu_0=\mu_\alpha+\mu_\beta$ lies within a gap separating two
bands. The existence of two (direct and indirect) gaps makes the
CL case principally different from 2L model.

One could expect that in the limit $t_{\|}\gg t_\perp$ the
properties of the system will be close to those of two-orbital
Hubbard chain\cite{Yama,Uima} because the orbital degeneracy is
nearly removed by strong longitudinal hopping. This is not the case,
however. The gap in the spectrum prevents formation of $SU(4)$
manifold formed by spin and orbital degrees of freedom. However the
interband transitions as well as the edge singularities in the DoS
influence the thermodynamic and optical properties of CL

In the remainder of this section we concentrate on the case
$\mu=0$ and consider the manifestation of singularities in the
excitation spectrum in different response functions of CL. The
density of states $\rho(E) =(2\pi)^{-1}\int dk
\delta(E-\varepsilon(k))$ is divergent at $E=\pm t_\perp$ and at
$E=\pm (t_\| +\sqrt{t_\perp^2 + t_\|^2})$. It  behaves as
\begin{eqnarray}
    \rho(E) &\simeq&  \Delta_{ind}^{-1/2}
    \sum_\pm \frac{\vartheta(-\Delta_{ind}\pm E)}{\sqrt{\pm E -\Delta_{ind}}}
\end{eqnarray}
at small energies $|E| \sim \Delta_{ind}$; here $\vartheta(x)=1$
at $x>0$. Such character of the density of states (see inset in
Fig. \ref{fig:f4}) should lead to observable anomalies in the
thermodynamic quantities. The specific heat $C$, compressibility
$\chi$ and dc conductivity $\sigma_{dc}$ are sensitive to indirect
gap. Setting all fundamental constants to unity we write for
thermodynamic quantities per unit cell
\begin{eqnarray}
   C & = & \int \frac{E^2 dE}{4T^2 \cosh^2 (E/2T)}     \rho(E) \nonumber \\
     &\sim & \left(\frac{T}{ \Delta_{ind}} \right)^{1/2},
     \quad T\gg\Delta_{ind} \nonumber \\
     &\sim &\left(\frac{ \Delta_{ind}}{T}
     \right)^{3/2} e^{- \Delta_{ind}/T},
     \quad T\ll\Delta_{ind}
     \end{eqnarray}
for the specific heat and
     \begin{eqnarray}
     \chi & = & \int \frac{dE}{4T \cosh^2 (E/2T)} \rho(E) \nonumber\\
     &\sim & \left({T}{ \Delta_{ind}} \right)^{-1/2},
     \quad T\gg\Delta_{ind} \nonumber\\
     &\sim &\left( \Delta_{ind} T \right)^{-1/2}
     e^{- \Delta_{ind}/T},
     \quad T\ll\Delta_{ind}
\end{eqnarray}
for the compressibility along the ladder. Assuming the elastic
quasiparticle lifetime $\tau_{el}$, the dc conductivity is given by
\begin{eqnarray}
  \sigma_{dc} & = & \tau_{el}
  \int \frac{dk}{4T \cosh^2 (\varepsilon_\pm (k)/2T)}
  \left(\frac{d\varepsilon_\pm(k)}{dk}\right)^2  \nonumber \\
     &\sim & \tau_{el} \sqrt{T \Delta_{ind}}
     ,\quad T\gg\Delta_{ind} \nonumber\\
     &\sim & \tau_{el} \sqrt{T \Delta_{ind}}
     e^{- \Delta_{ind}/T},
     \quad T\ll\Delta_{ind}
\end{eqnarray}

In contrast to behavior of thermodynamic quantities sensitive to
indirect gap, the optical conductivity response is sensitive to
direct gap. It is non-zero at $|\omega|>\Delta$ and is given by
\cite{ArZey}:
\begin{eqnarray}
  \sigma_{opt}(\omega) & = & \frac{1}{\omega}
  \int dk [n(\varepsilon_-(k)) - n(\varepsilon_+(k))]
   \nonumber\\ &&\times
  \frac{t_\perp^2 t_\|^2 \sin^2 k}{t_\perp^2 + t_\|^2 \cos^2 k}
  \delta(\omega +\varepsilon_-(k) - \varepsilon_+(k))
  \nonumber \\
     &\sim &
     \frac{t_\|\Delta^2 }{\omega^2 \sqrt{\omega^2 - \Delta^2}}
     ,\quad T\ll\Delta \nonumber \\
     &\sim &
     \frac{t_\|\Delta^2 }{|\omega| T \sqrt{\omega^2 - \Delta^2}}
     ,\quad T\agt \Delta
\end{eqnarray}
We thus see a continuous absorption band in $\sigma_{opt}(\omega)$
in contrast to the case of 2L ladder, where the optical
conductivity consists of a single line, $\sigma_{opt}(\omega)
\propto \delta(\omega \pm \Delta)$.

Interband optical transitions may result in creation of excitons
with a center-of-mass momentum $Q=\pi$. Unlike the case of weak
longitudinal hopping (section II), these excitons arise in the
conventional way due to attractive Coulomb interaction between
electrons and holes. The bound electron-hole states formed on the
almost flat parts of the spectrum over the indirect gap are
characterized by heavy effective mass. We therefore expect that
these excitons are nearly localized and one may neglect the
processes of their coherent motion.

\section{Conclusions and perspectives}

In this paper we have demonstrated several fragments of the rich
phase diagram for centipede ladder. It was shown in our previous
studies \cite{kak05a,kak05b} that the half-filled CL is an example
of spin 1 chain with soft triplet-singlet excitation, which does not
belong to the Haldane gap universality class. Now we have found that
the 1/4 filled CL also demonstrate unconventional properties both in
spin and charge sectors.

We considered here only three limiting cases of the quarter filled
CL where the character of charge and spin excitations may be
revealed by means of relatively simple arguments. First, we have
found the regime where CL behaves as a spin 1 chain with
ferromagnetic effective longitudinal exchange and magnon-like
excitation spectrum (Section II). It is shown that the system is
unstable against dimerization at strong enough on-rung ferromagnetic
exchange. The only coherent mode in the charge sector is the gapped
triplet charge-transfer exciton.

One should stress that the dimerisation effect offered in this
section has no analogs in current literature. Usually the
dimerization  mechanism in spin chains is either explicitly built in
the Heisenberg Hamiltonian or arises as a result of competition
between the nearest and next nearest exchange coupling \cite{Micol}.
As to the two-leg ladders, the dimerized configurations were
considered in a context of RVB states (rung-dimer states vs
leg-dimer states)\cite{Naka}. In our case the lattice period is
doubled as a result of competition between the {\sl on-rung} hopping
and exchange coupling. Experimentally, rung dimerization as a result
of charge redistribution between legs and rungs was detected
optically\cite{exper1} in two-leg compounds AV$_6$O$_{15}$
(A=Na,Sr). However, in vanadium bronzes this effect is driven mainly
by the valence instability of V ions.

Second, we have described the situation, where the vacancies (empty
rungs) and spin singlets (doubly occupied rungs) may propagate
either as coherent excitons or as independent spinless fermionic
quasiparticles with different Fermi velocities (Section III). The
dispersion law is gapped in the former case and gapless in the
second case, where the charge sector may be described as a
two-component Fermi liquid with weak short-range interaction. The
spin subsystem behaves as the RVB-type spin liquid, where the
spinons propagate in presence of spin 1 and dangling bond defects.

Third, we have considered the limit of strong longitudinal hopping,
where the CL behaves as a specific version of two-orbital Hubbard
chain. The peculiar features of this limit is the appearance of a
gap in the half-occupied lower Hubbard band. This gap has purely
hybridization nature, but the van-Hove singularities in the density
of states strongly influence the thermodynamic and optical
properties of CL.

Our analysis of these three cases by no means exhausts the variety
of unusual properties of CL at partial filling. In particular, we
didn't include in our Hamiltonian direct longitudinal exchange which
can enhance the indirect kinetic exchange or compete with it. We
didn't consider the phase transitions between different phases,
which may have their own peculiarities as compared to currently
studied quantum phase transitions in low-dimensional systems
\cite{sash}. We also didn't discuss possible lattice distortions,
which may accompany dimerization. All these and many other open
questions are left for future investigation.

The centipede ladder shown in Fig. \ref{fig:f1} is the simplest
example of the family of quasi 1D systems intermediate between
chains and two-leg ladders. The experimental realization of such
system (organic biradical crystals PNNNO \cite{exp}) is already
known, but one may easily imagine molecular chains decorated not
with rungs but with radicals forming closed loops, zigzags etc. We
believe that the study of excitation spectra,  magnetic and
transport properties of this family will bring new unexpected
results.

\section{Acknowledgments}
We are grateful to F.Assaad, B.N.Narozhny and A.M.Tsvelik for useful
and stimulating discussions. DNA thanks ICTP for the hospitality.
MNK appreciates support from the Heisenberg program of the DFG and
the SFB-410 research grant during initial stage of the CL project.
MNK also acknowledges support from U.S. DOE, Office of Science,
under Contract No. W-31-109-ENG-39.

\appendix*
\section{Calculation of the superexchange interaction}

In order to calculate $J_\|$ we note that two spins $S=1$ at even
sites in Eq.\ (\ref{FMeffmodel}) and in Fig.\  \ref{fig:f1b} can be
combined into total spin $S_{tot}=0, 1, 2$. The scalar product ${\bf
S}_i \cdot {\bf S}_{i+2}=\frac{1}{2}(({\bf S}_i+{\bf
S}_{i+2})^2-{\bf S}_i^2-{\bf S}_{i+2}^2)$ in (16) attains values
$-2,-1, 1$ for these values $S_{tot}$, respectively. The value of
$3J_\|$ can hence be naively calculated as a difference in
fourth-order corrections (\ref{h4}) to the states $S_{tot}=0$ and
$S_{tot}=2$.

Let us use a shorthand notation $|m, m'\rangle$ for a quantum state
$|Tm, i ; Tm',i+2 \rangle$, and $\Phi_{L,m}$ for the state with
total spin $S_{tot}=L$ and its $z-$component $m$. The weight of
$\Phi_{L,m''}$ in $|m, m'\rangle$ is the Clebsch-Gordan coefficient
$C^{Lm''}_{1m1m'}$. Then the states we are interested in,
$S_{tot}=0$ and $S_{tot}=2$, are given by $ \Phi_{0,0} = ( |-1,1
\rangle + |1,-1 \rangle -|0,0 \rangle )/\sqrt{3}$ and $ \Phi_{2,2}
|1,1 \rangle $.

Further, we denote the combinations  $|+\sigma_1,i;
+\sigma_3,i+2\rangle $  in $|\phi^{(2)}\rangle$, Eq.\ (\ref{virt})
as "non-local triplet" states, e.g.\  $|+\uparrow,i;
+\uparrow,i+2\rangle  = |\widetilde{T1}, i+1\rangle$ etc. The
intermediate state in Fig.\ 3, obtained after two longitudinal hops
in terms of its spin components, is rather obviously classified in
the same terms of total spin, composed of two triplets, $|{T}m,
i+1\rangle$ and $|\widetilde{Tm}, i+1\rangle$. Similarly denoting
$|\widetilde{Tm}, {i+1} ;  Tm',i+1 \rangle  = |\widetilde{m,
m'}\rangle $ we have $ \widetilde \Phi_{0,0} = ( |\widetilde{-1,1}
\rangle + |\widetilde{1,-1} \rangle  - |\widetilde{0,0} \rangle
)/\sqrt{3}$ etc.

It then can be shown that (projecting at each step at the states
with lowest energy, bonding and triplet, as discussed above)
\begin{widetext}
 \begin{eqnarray}
 &&{\cal H}_{l}^2 |-1,1 \rangle
 = \frac{t_\|^2}{2\sqrt 2}
 |+\downarrow,i;T0, i+1 ; +\uparrow,i+2\rangle\nonumber\\
 &&{\cal H}_{l}^2 (|-1,1 \rangle  +  |1,-1 \rangle )  = \frac{t_\|^2}2
|\widetilde{0,0} \rangle = \frac{t_\|^2}2 [\sqrt{\frac23}\widetilde
\Phi_{2,0} -
 \sqrt{\frac13}\widetilde \Phi_{0,0} ]
\nonumber \\
 &&{\cal H}_{l}^2 |0,0 \rangle  = \frac{t_\|^2}4
 ( |\widetilde{-1,1} \rangle +
|\widetilde{1,-1} \rangle  + |\widetilde{0,0} \rangle )=
\frac{t_\|^2}4 [2\sqrt{\frac23}\widetilde \Phi_{2,0} +
 \sqrt{\frac13}\widetilde \Phi_{0,0} ]
  \end{eqnarray}
\end{widetext}
which leads to ${\cal H}_{l}^2 \Phi_{0,0} = - \frac {t_\|^2}4
\widetilde \Phi_{0,0}$. Similarly,  one obtains ${\cal H}_{l}^2
\Phi_{2,2} = \frac {t_\|^2}{2}  \widetilde \Phi_{2,2}$. Because the
projected action of ${\cal H}_{l}^2$ is self-conjugate, we obtain
${\cal H}_{l}^4 \Phi_{0,0} = \frac {t_\|^4}{16} \Phi_{0,0}$ and
${\cal H}_{l}^4 \Phi_{2,2} = \frac {t_\|^4}{4} \Phi_{2,2}$.

We would also like to exclude more complex structure of the
effective Hamiltonian, namely
\begin{equation}
{\cal H}_{eff}= -J_\|\sum_i ({\bf S}_i \cdot {\bf S}_{i+2}  + c_1
({\bf S}_i \cdot {\bf S}_{i+2} )^2)
\end{equation}
which is not prohibited by symmetry.

To this end, we calculate also the correction to states $\Phi_{1,m}$
of total spin $L=1$. Within the same logic, we have ${\cal H}_{l}^2
(|1,-1 \rangle  - |-1,1 \rangle) = {\cal H}_{l}^2
\sqrt{2}\Phi_{1,0}= \frac{t_\|^2}2 |\widetilde{S}, {i+1} ;  T0,i+1
\rangle$ where $\widetilde{S}$ non-local singlet.  Therefore $ {\cal
H}_{l}^4 \Phi_{1,0} = \frac{1}{8}t_\|^4 \Phi_{1,0}$ and for our
values $L=0,1,2$ we have the corrections $-1/16, -1/8, -1/4$ in
units of $t^4_\|/\Delta_T^3$, respectively. It means that $J_\| =
{t^4_\|}/({16\Delta_T^3})$ and  $\quad c_1 =0$. Thus the anisotropy
term $({\bf S}_i \cdot {\bf S}_{i+2} )^2$ does not appear in the
forth order of longitudinal hopping. The reason of it is very
simple: in order to get the four-spin $S=1$ interaction within the
scheme shown on Fig.\  \ref{fig:f1b} where four spins $s=1/2$
participate, one needs at least $8$ processes of longitudinal
hopping.


\end{document}